\documentclass[10pt,a4paper]{article}
\usepackage[english]{babel}
\usepackage[latin1]{inputenc}
\usepackage{amsfonts,amsbsy,bm,euscript,mathrsfs}
\usepackage{amssymb,stmaryrd,faktor}
\usepackage[tbtags]{amsmath}
\usepackage{bbm}
\usepackage{graphicx}
\usepackage[title,titletoc]{appendix}
\usepackage[bookmarks=true,colorlinks=true,linkcolor=blue,citecolor=blue,urlcolor=blue,bookmarksnumbered]{hyperref}

\usepackage{dsfont}
\usepackage{collref}
\usepackage{lmodern}
\usepackage{mathrsfs}
\usepackage{mathtools}
\usepackage{bbm}
\usepackage{braket}
\usepackage{slashed}

\usepackage{graphicx}
\usepackage{booktabs}
\usepackage{subfig}
\usepackage{tikz}
\usetikzlibrary{plotmarks,calc,decorations,decorations.pathmorphing}

\numberwithin{equation}{section}

\makeatletter
\renewcommand\section{\@startsection {section}{1}{\z@}
{-3.5ex \@plus -1ex \@minus -.2ex}
{2.3ex \@plus.2ex}
{\normalfont\Large\bfseries}}
\renewcommand\subsection{\@startsection{subsection}{2}{\z@}
{-3.25ex\@plus -1ex \@minus -.2ex}
{1.5ex \@plus.2ex}
{\normalfont\large\bfseries}}
\makeatother

\newcommand{\alg}[1]{\mathfrak{#1}}

\newcommand{\acomm}[2]{\{#1,#2\}}

\newcommand{\matId}{\mathfrak{1}}

\def\tr{{\rm tr}}

\def\bE{\mathsf{E}}

\usepackage{blkarray}

\begin{document}

\thispagestyle{empty}
\begin{flushright}\footnotesize\ttfamily
DMUS-MP-17/08
\end{flushright}
\vspace{2em}

\begin{center}

{\Large\bf \vspace{0.2cm}
{\color{black} \large On $AdS_2/CFT_1$ transfer matrices, Bethe ansatz and scale invariance}} 
\vspace{1.5cm}

\textrm{Alessandro Torrielli\footnote{\texttt{a.torrielli@surrey.ac.uk}}}

\vspace{2em}

\vspace{1em}
\begingroup\itshape
Department of Mathematics, University of Surrey, Guildford, GU2 7XH, UK
\par\endgroup

\end{center}

\vspace{2em}

\begin{abstract}\noindent 

\end{abstract}

We explicitly calculate the $AdS_2 \times S^2 \times T^6$ transfer-matrix eigenvalues in the massless sector using the exact integrable S-matrix, for up to 5 particles. This enables us to conjecture the general pattern. We use the conjectured form of the eigenvalues to write down a set of massless Bethe ansatz equations. The same procedure applies to the relativistic as well as to the non-relativistic situation. In the relativistic case, the right and left modes decouple. We speculate that the relativistic massless Bethe ansatz we obtain in that case might capture the integrable structure of an underlying 2D critical theory. We finally take advantage of some remarkable simplifications to make progress in the massive case as well. 

\newpage

\overfullrule=0pt
\parskip=2pt
\parindent=12pt
\headheight=0.0in \headsep=0.0in \topmargin=0.0in \oddsidemargin=0in

\vspace{-3cm}
\thispagestyle{empty}
\vspace{-1cm}

\tableofcontents

\setcounter{footnote}{0}

\section{\label{sec:Intro}Introduction}

AdS/CFT integrability \cite{rev} provides a powerful testing ground for our general ideas regarding holography, whose physical manifestation assumes a rich variety of forms as one varies in dimension \cite{sym}. One of the lowest dimension one can reach involves strings on an  $AdS_2 \times S^2 \times T^6$ background \cite{ads2}. The holographic description is anticipated to either be a superconformal quantum mechanics, or a chiral two-dimensional CFT \cite{dual,gen}. The model
is based on a Metsaev-Tseytlin action \cite{Metsaev:1998it} on the $\frac{PSU(1,1|2)}{ SO(1,1) \times SO(2)}$ supercoset. Classical integrability (cf. \cite{Bena:2003wd}) has been explicitly shown up to quadratic order in the fermions \cite{Sorokin:2011rr,Cagnazzo:2011at}.

In \cite{Hoare:2014kma}, an exact S-matrix was obtained by postulating a centrally-extended $\mathfrak{psu}(1|1)^2$ symmetry superalgebra acting on excitations above the BMN vacuum \cite{Berenstein:2002jq,amsw}, along the lines of \cite{beis0}. The S-matrix for massive modes satisfies crossing and unitarity, but the dressing factor is still unknown. Under reasonable assumptions for such dressing factor, the near-BMN expansion agrees with perturbation theory \cite{amsw}. The magnon representations in $AdS_2$ are {\it long} for any non-zero mass. The massless representations are instead {\it short}, and the S-matrix is obtained as an appropriate limit of the massive one, following the general prescription \cite{Zamol2,BogdanLatest}, for all combinations of right and left movers. The scattering theory displays Yangian symmetry \cite{Hoare:2014kma,Hoare:2014kmaa}. Perturbation theory is much trickier in the presence of massless modes \cite{amsw,MI}, and it remains to be fully understood. Although we will not have anything to add about perturbative issues in this paper, the long-term goal of understanding the massless sector is largely driving this investigation. 

Massless scattering plays a very special role in the theory of integrable systems. It was introduced by Zamolodchikov as a method of describing certain examples of renormalisation group flow between critical points \cite{Zamol2}. The properties of massless scattering are different from those of massive S-matrices - see {\it e.g.} \cite{Borsato:2016xns}. In string theory, the absence of relativistic invariance after gauge fixing constitutes an additional complication \cite{BogdanLatest}. It is therefore worthwhile to first study the relativistic limit, where one can make contact with known results in ${\cal{N}}=1$ supersymmetric theories \cite{Fendley:1990cy}. For scattering matrices between only right-right and left-left movers, the relativistic limit is non-trivial. This was the approach taken in \cite{Fontanella:2017rvu} to tackle the problem.

Even in the massless relativistic limit, the S-matrix still has the maximal number of non-zero entries, much like the XYZ or the eight-vertex model \cite{Baxter:1972hz}, and not unlike other instances of ${\cal{N}}=1$ supersymmetric theories \cite{Schoutens,MC}. The associated transfer matrices do not admit a reference state, invalidating the use of the algebraic Bethe ansatz \cite{Levkovich-Maslyuk:2016kfv}. This makes it difficult to test, for instance, the proposal of \cite{Sorokin:2011rr} for the massive Bethe equations, starting directly from the S-matrix. A variety of methods exist to circumvent this obstacle, besides Baxter's original approach of functional relations; see for instance \cite{Nepo}. In the $AdS_2$ case it is possible \cite{Fontanella:2017rvu} to rely on the so-called {\it free-fermion condition}, and to follow the same approach as in \cite{MC,Ahn:1993qa} to reduce the problem to solving a particular {\it inversion relation} \cite{Zamolodchikov:1991vh}. 

The free-fermion condition is valid more generally for massive $AdS_2$ scattering as well \cite{Hoare:2014kma}, and it corresponds to a $\alg{u}(1)$ symmetry of the string theory \cite{amsw}. Thanks to this symmetry, one can in principle find a reference state for the {\it complete} S-matrix, which is made out of two copies of the centrally-extended one. This echoes a  technique which was perfected for particular models in \cite{Faddeev:1995nf}. It seems however to be technically prohibitive to proceed along that route in the $AdS_2$ problem. The strategy of \cite{MC,Ahn:1993qa} turns out to be simpler to implement in this case, as we will shortly review. 

As a result of this procedure, \cite{Fontanella:2017rvu} derived an inversion relation for the transfer-matrix eigenvalues and a set of auxiliary Bethe equations, identifying the location of the eigenvalues' potential zeroes. The inversion relation could not be solved purely on the basis of the information about zeroes and poles, as the asymptotic behaviour at infinity of the eigenvalues displays essential singularities. This is the reason why we have embarked in this paper on a brute-force evaluation of the eigenvalues for a small number of particles. We will explicitly calculate the transfer-matrix eigenvalues up to 5 particles. This will provide us with some confidence in conjecturing the general pattern. We shall then use the conjectured form of the eigenvalues to write down the momentum-carrying massless Bethe equations. Since the same procedure applies to the relativistic as well as to the non-relativistic situation, it is relatively straightforward to encompass both cases at a the same time. However, it is only in the relativistic case that the right and left modes decouple. Following Zamolodchikov, we speculate that this should describe a critical fixed point. Therefore, we are brought to conclude that the relativistic massless Bethe ansatz we propose - which is completely non-perturbative - might capture the integrable structure underlying some limiting 2D critical field theory, which is a worldsheet theory\footnote{We thank Diego Bombardelli, Bogdan Stefa\'nski and Roberto Tateo for many clarifying discussions about this point.}. The roots of this phenomenon were originally noticed in \cite{Borsato:2016xns} in the $AdS_3$ context, and will be treated in that situation with considerably more detail in the upcoming paper \cite{Diego}.   

In the last part of the paper, we demonstrate that the very same procedure is applicable to the massive case as well. This is thanks to some astonishing simplifications, in spite of the great complication of the massive S-matrix entries. Although it is still tremendously difficult to reduce the outcome of the diagonalisation process to a manageable expression, which could be for instance compared to the proposal of \cite{Sorokin:2011rr}, this makes us hope that we could eventually solve the massive case as well with a greater effort along these very same lines.

\section{\label{sec:Aux}Auxiliary massless Bethe equations}

In this section, we summarise the analysis of \cite{Fontanella:2017rvu} for the auxiliary set of $AdS_2$ Bethe equations. 

\subsection{\label{sec:AuxRel}Relativistic}

The lack of a reference state prevents the application of the algebraic Bethe ansatz. The method relying on the {\it free-fermion condition} \cite{Felderhof,MC,Ahn:1993qa} allows one to solve this problem. 

The first step is to switch from the R-matrix to the S-matrix, and then proceed to ignore any further fermionic sign. The S-matrix is given by  
\begin{eqnarray}
S = \begin{pmatrix}A&B\\C&D\end{pmatrix},
\end{eqnarray}
where the matrix acts on the first ({\it auxiliary}) space, while $A$, $B$, $C$ and $D$ act on the second ({\it quantum}) space as
\begin{eqnarray}
A = a_+ \, \bE_{11} \, + \, b_+ \, \bE_{22}, \qquad B = d_+ \, \bE_{12} \, + \, c_- \, \bE_{21}, \qquad C = c_+ \, \bE_{12} \, + \, d_- \, \bE_{21}, \qquad D = b_- \, \bE_{11} \, + \, a_- \, \bE_{22}, \nonumber 
\end{eqnarray}
in terms of the standard basis $\bE_{ij}$ of $2 \times 2$ matrices. One has, for the S-matrix dubbed {\it solution 3} in \cite{Fontanella:2017rvu},
\begin{eqnarray}
a_+ = a_- = 1, \qquad b_- = - b_+ = 1, \qquad c_+ = c_- = e^{\frac{\theta_2 - \theta_1}{2}}, \qquad d_+ = - d_- = e^{\frac{\theta_2 - \theta_1}{2}}
\end{eqnarray} 
(momentarily suppressing the dressing factor, which is reinstated at the end), satisfying the {\it free-fermion condition}:
\begin{eqnarray}
a_+ a_- + b_+ b_- = c_+ c_- + d_+ d_-
\end{eqnarray}
- cf. also \cite{Bazhanov:1981eg,Mitev:2012vt}.
The monodromy matrix $M$ and the transfer matrix $T$ are defined as
\begin{eqnarray}
M =  S_{01} (\theta_0 - \theta_1) ... S_{0N} (\theta_0 - \theta_N), \qquad T = \tr_0 M,
\end{eqnarray}
with $S_{0i}$ acting in spaces $0$ ({\it auxiliary space}) and $i$ ({\it $i$-th quantum space}). 

We now define a new S-matrix $S^{(1)}$ by the replacement
\begin{eqnarray}
&&a_\pm \to a^{(1)}_\pm = - b_\pm, \qquad b_\pm \to b^{(1)}_\pm = a_\pm, \qquad c_\pm \to c^{(1)}_\pm = c_\pm, \qquad d_\pm \to d^{(1)}_\pm = - d_\pm.
\end{eqnarray}  
$S^{(1)}$ still satisfies the free-fermion condition. Then, we write
\begin{eqnarray}
\label{traccia}
T \, T^{(1)} &=& \tr_0 \Big[ S_{01} (\theta_0 - \theta_1) ... S_{0N} (\theta_0 - \theta_N) \big]\times  \tr_{0'} \Big[S^{(1)}_{0'1} (\theta_0 - \theta_1) ... S^{(1)}_{0'N} (\theta_0 - \theta_N)\Big] \nonumber \\
&=& \tr_{0 \otimes 0'} \prod_{i=1}^N S_{0i} (\theta_0 - \theta_i) \otimes S^{(1)}_{0'i}(\theta_0 - \theta_i),  
\end{eqnarray} 
where the tensor product is over the two auxiliary spaces $0$ and $0'$ associated with $T$ and $T^{(1)}$, respectively, with a common auxiliary variable $\theta_0$. There now exists a similarity transformation on $S_{0i} (\theta_0 - \theta_i) \otimes S^{(1)}_{0'i}(\theta_0 - \theta_i)$ which puts it into upper triangular form. Such a transformation is performed at each site, but it is independent of the inhomogeneities $\theta_i$, hence it telescopically cancels in (\ref{traccia}), rendering straightforward the task of taking the trace. The matrix
\begin{eqnarray}
\label{x}
X = \frac{1}{\sqrt{2}} \begin{pmatrix}0&1&1&0\\1&0&0&1\\1&0&0&-1\\0&1&-1&0\end{pmatrix} = X^{-1}
\end{eqnarray}
achieves in fact
\begin{eqnarray}
\label{suchthatx}
X S_{0i} \otimes S^{(1)}_{0'i} X^{-1} = X \begin{pmatrix}A A^{(1)}&A B^{(1)}&B A^{(1)}&B B^{(1)}\\A C^{(1)}&A D^{(1)}&B C^{(1)}&B D^{(1)}\\C A^{(1)}&C B^{(1)}&D A^{(1)}&D B^{(1)}\\C C^{(1)}&C D^{(1)}&D C^{(1)}&D D^{(1)}\end{pmatrix} X^{-1} = \begin{pmatrix} m_+ & * & * & *\\0 & n_+ & * & * \\ 0 & 0 & n_- & * \\ 0 & 0 & 0 & m_-\end{pmatrix},
\end{eqnarray}
where
\begin{eqnarray}
&&m_\pm = \big(1\pm e^{\theta_i - \theta_0}\big) \mathfrak{1}, \qquad n_\pm =  \big(1\pm e^{\theta_i - \theta_0}\big) \sigma_3, \qquad \mathfrak{1} = \bE_{11} + \bE_{22}, \qquad \sigma_3 = \bE_{11} - \bE_{22}.
\end{eqnarray}
Since $\tr_{0 \otimes 0'} = \tr_4$, one finds
\begin{eqnarray}
\label{fromm}
&&T T^{(1)} = (1+F)
\times \Bigg[\prod_{i=1}^N \Big(1 + e^{-(\theta_0 - \theta_i)}\Big) +   \prod_{i=1}^N \Big(1 - e^{-(\theta_0 - \theta_i)}\Big)\Bigg],
\end{eqnarray} 
where $F$ denotes the fermionic number of the transfer-matrix eigenstate ($F=1$ for bosonic eigenstates, $F=-1$ for fermionic ones). 

We then use that
\begin{eqnarray}
\label{rela}
S_{0'i}^{(1)} = \tau \, \sigma_1 \, S_{0'i}(\theta_0 - \theta_i + i \pi) \sigma_1^{-1} \, \tau^{-1}, \qquad \sigma_1 = \bE_{12} + \bE_{21}, \qquad \tau = \bE_{11} + i \bE_{22},
\end{eqnarray}
where the similarity transformation acts on the space $0'$. Formula (\ref{rela}) also confirms that, in particular, the S-matrix $S^{(1)}$ satisfies the Yang-Baxter equation. We obtain, as a consequence of (\ref{rela}), that 
\begin{eqnarray}
\label{thats}
T^{(1)} (\theta_0) = T(\theta_0 + i \pi). 
\end{eqnarray}
By virtue of this, (\ref{fromm}) takes the form of a specific {\it inversion relation}\footnote{The use of such inversion relations originated with Baxter \cite{BaxBook}. It is a very interesting question how unique the procedure is (modulo ``gauge" equivalences) to upper-triangularise tensor products of matrices, whether or not one relies on transforming the functional arguments, or the Yang-Baxter equation. Ref. \cite{Felderhof} exhaustively discusses the XYZ model. Besides \cite{BaxBook} (cf. section 14.3), we have not found a more general treatment. We thank B. Stefa\'nski for raising the issue.}. The final task is reduced to a factorisation problem. 

It is relatively easy to show that the eigenvalues of the transfer matrix are $2 \pi i$-periodic and meromorphic in $\theta_0$. Hence, it would in principle be sufficient to know the location of their zeroes and poles in the strip $\theta_0 \in [-\pi, \pi)$ to uniquely fix them, provided one had suitable asymptotics at infinity. In this case, it is immediate to see from (\ref{fromm}) that potential zeroes can arise when
\begin{equation}
\label{sub}
\prod_{i=1}^N \tanh \frac{\theta_0 - \theta_i}{2} = -1, \quad k=1,...,M.
\end{equation}
Eq.s (\ref{sub}) play the role of {\it auxiliary} Bethe equations in the Bethe ansatz. As we will described later in the paper, the {\it main} (momentum-carrying) Bethe equations involve the transfer-matrix eigenvalues themselves, on which one imposes Bethe's familiar periodicity condition. However, the task is now to separate the zeroes of $T$ from those of $T^{(1)}$ (tantamount to solving the inversion relation), and this is made difficult by the very asymptotics of the eigenvalues, which either vanish or have an essential singularity at infinity. Factoring out the singular term and going to a reduced transfer matrix, as in \cite{Zamolodchikov:1991vh}, does not seem to be straightforward either.

\subsection{\label{sec:AuxNonRel}Non-relativistic}

As shown in \cite{Fontanella:2017rvu} by direct inspection of the respective S-matrices, the {\it non-relativistic} case simply turns out to be obtained by replacing
\begin{eqnarray}
\label{repl}
&&e^{\frac{\theta_0 - \theta_i}{2}} \to \sqrt{\frac{\tan \frac{p_0}{4}}{\tan \frac{p_i}{4}}}, \qquad 1 \pm e^{\theta_0 - \theta_i}\to 1 \pm \frac{\tan \frac{p_0}{4}}{\tan \frac{p_i}{4}},
\end{eqnarray} 
where the map $\theta_0 \to \theta_0 + i \pi$ is replaced by $p_0\to -p_0$. The auxiliary Bethe equations (\ref{sub}) get immediately replaced by
\begin{equation}
\label{just}
\prod_{i=1}^N \frac{\sin \frac{q_k + p_i}{4}}{\sin \frac{q_k - p_i}{4}} = - 1, \qquad k=1,...,M.
\end{equation}

This represented a partial success of \cite{Fontanella:2017rvu} in structurally matching the naive massless limit of the Bethe ansatz, which was proposed in \cite{Sorokin:2011rr} for the massive modes. Nevertheless, the main Bethe equations were not addressed in this case either, just as in the relativistic situation of the previous section, given the difficulty which persists in solving the corresponding inversion relation. 

\section{\label{sec:Main}Momentum-carrying massless Bethe equations}

In this section, we attempt to complete the task begun in \cite{Fontanella:2017rvu}, and provide the momentum-carrying Bethe equations for massless $AdS_2$ integrable scattering, to supplement the set of auxiliary ones which were found there. We will first describe the relativistic situation, and then generalise the analysis to the non-relativistic one, which is seen to proceed along very similar lines.

\subsection{\label{sec:MainRel}Relativistic} 

The main obstacle \cite{Fontanella:2017rvu} encountered in the derivation of the transfer-matrix eigenvalues, even knowing its set of potential zeroes (auxiliary Bethe equations), was due to the large-$\theta$ asymptotics, displaying either zeroes or essential singularities. What we will do here is to calculate the eigenvalues for a small number of particles, namely up to 5, and recognise the location of the zeroes. Going up to 5 particles will give us enough confidence in spotting the general trend. This will allow to recast the eigenvalue in a suggestive form, which we will then extrapolate to arbitrary number of particles as part of our conjecture.   

Let us begin by writing the S-matrix - solution 3 of \cite{Fontanella:2017rvu} - as
\begin{equation}
\label{S}
S (\theta_{12}) \equiv S (\theta_1-\theta_2) = \widehat{\mathfrak{1}} + g(\theta_1 - \theta_2) \, \widehat{\mathbb{P}},\qquad g(\theta) \equiv e^{- \frac{\theta}{2}}, \qquad \theta_{ij} \equiv \theta_i - \theta_j,
\end{equation}
where
\begin{eqnarray}
&&\widehat{\mathfrak{1}} \equiv \bE_{11} \otimes \bE_{11} - \bE_{11} \otimes \bE_{22} + \bE_{22} \otimes \bE_{11} + \bE_{22} \otimes \bE_{22},\nonumber\\
&&\widehat{\mathbb{P}} \equiv \bE_{12} \otimes \bE_{12} + \bE_{12} \otimes \bE_{21} + \bE_{21} \otimes \bE_{12} - \bE_{21} \otimes \bE_{21}.
\end{eqnarray}
The index 1 corresponds to the bosonic state $|\phi\rangle$, and 2 to the fermionic state $|\psi\rangle$. We have omitted the dressing factor, which is straightforward to reinstate at the very end. 
 
We define a rescaled transfer matrix as
\begin{equation}
\label{T}
T_N = \frac{1}{2} \, \tr \, S_{01}(\theta_{01}) \, S_{02}(\theta_{02}) \, ... \, S_{0N}(\theta_{0N}), \qquad \theta_{0i} = \theta_0 - \theta_i, \quad i=1,...,N.
\end{equation}
Recall that we denote by $\theta_0$ the variable associated to the auxiliary space. The factor of $\frac{1}{2}$ in (\ref{T}) is for later convenience: because a systematic factor $2$ does appears in all the eigenvalues as a result of the calculation, we momentarily eliminate it to achieve a lighter notation, and reinstate it only at the very end when writing the Bethe equations (which will therefore involve twice the eigenvalue of $T_N$).

What follows is a series of technical sections on the explicit diagonalisation procedure. In case the reader were interested in seeing the final conjecture, they could skip the details in the first instance if they preferred to, and resume at section \ref{sec:Conj}.

\subsubsection{\label{sec:1p}2 particles}

We begin our exploration of small numbers of particles with the case $N=1$, {\it i.e.} 2 particles including the auxiliary one. This is the smallest value of $N$ for which one can have any scattering at all.

This case is of course absolutely straightforward, and reported here only for completeness. Given that 
\begin{equation}
\tr \, \bE_{ij} = \delta_{ij},
\end{equation}
one has
\begin{equation}
T_1 = \bE_{11}
\end{equation}
which annihilates the fermionic state $|\psi\rangle$, and whose bosonic action is diagonalised by 
\begin{equation}
T_1 |\phi\rangle = |\phi\rangle.
\end{equation}
The eigenvalue is $1$, and the eigenvector does not depend on $\theta_0$. This is naturally dictated by the fact that the transfer matrix generates the integrable charges in involution, as it commutes with itself at different values of the spectral parameter (auxiliary variable) $\theta_0$ by virtue of the RTT relations. Hence, it can be diagonalised simultaneously at different values of $\theta_0$, therefore its eigenstates do not depend on $\theta_0$. This will of course always be the case for any $N$, as we will explicitly verify below.

The auxiliary Bethe equation simply reads 
\begin{equation}
\tanh \frac{\beta - \theta_1}{2} = -1,
\end{equation}
which only has $\beta = - \infty$ as a solution. The eigenvalue clearly has no zeroes.

\subsubsection{\label{sec:2p}3 particles}

Let us consider now $N=2$. Recalling that we ignore fermionic signs, and using
\begin{equation}
\bE_{ij} \bE_{mn} = \delta_{jm} \bE_{in},
\end{equation}
one easily finds
\begin{equation}
T_2 = \bE_{11} \otimes \bE_{11} + \bE_{22} \otimes \bE_{22} + g(\theta_{01})g(\theta_{02}) \, (\bE_{12} \otimes \bE_{12} - \bE_{21} \otimes \bE_{21}).
\end{equation}
When acting on the quantum spaces $1$ and $2$, the operator $T_2$ annihilates fermionic states, and acts on bosons as follows:
\begin{equation*}
  T_2 = \mathfrak{1} + \, e^{-\theta_0}
  \begin{blockarray}{*{2}{c} l}
    \begin{block}{*{2}{c}>{$\footnotesize}l<{$}}
      |\phi\rangle \otimes |\phi\rangle & |\psi\rangle \otimes |\psi\rangle \\
    \end{block}
    \begin{block}{[*{2}{c}]>{}}
      0 & t_{12} & |\phi\rangle \otimes |\phi\rangle \\
      -t_{12} & 0 & |\psi\rangle \otimes |\psi\rangle  \\
    \end{block}
  \end{blockarray} \qquad \mbox{with} \, \, \, \, \, \, t_{ij} = e^\frac{\theta_i + \theta_j}{2},
\end{equation*}
$\mathfrak{1}$ being the identity matrix $\sum_{a=1}^2 \bE_{aa} \otimes \bE_{aa}$.

Clearly, the eigenstates do not depend on $\theta_0$ and are given by
\begin{equation}
|v_\pm\rangle = |\phi\rangle \otimes |\phi\rangle \pm i |\psi\rangle \otimes |\psi\rangle, \qquad \lambda_\pm = 1 \pm i e^{-\theta_0} \, t_{12},
\end{equation}
reported with their respective eigenvalues under $T_2$.

The auxiliary Bethe equations now read 
\begin{equation}
\tanh \frac{\beta - \theta_1}{2} \, \tanh \frac{\beta - \theta_2}{2} = -1,
\end{equation}
which has solutions
\begin{equation}
e^{\frac{\beta}{2}} = \pm e^{i \frac{\pi}{4}} e^\frac{\theta_1 + \theta_2}{4} \equiv x_\pm, \qquad e^{\frac{\beta}{2}} = \pm e^{3 i \frac{\pi}{4}} e^\frac{\theta_1 + \theta_2}{4} \equiv y_\pm.
\end{equation}
We see that
\begin{equation}
\lambda_- = e^{-\theta_0} \big(e^{\frac{\theta_0}{2}} - x_+\big)\big(e^{\frac{\theta_0}{2}} - x_-\big), \qquad \lambda_+ = e^{-\theta_0} \big(e^{\frac{\theta_0}{2}} - y_+\big)\big(e^{\frac{\theta_0}{2}} - y_-\big),
\end{equation}
which shows that the auxiliary Bethe roots determine the zeroes of the transfer-matrix eigenvalues.

\subsubsection{\label{sec:3p}4 particles}

The case of 4 particles , namely $N=3$, is slightly more involved, but rather illuminating.

Afted a tedious but straightforward calculation, we obtain
\begin{eqnarray}
T_3 &=& \bE_{11} \otimes \bE_{11} \otimes \bE_{11} + \bE_{11} \otimes \bE_{22} \otimes \bE_{22} + \bE_{22} \otimes \bE_{11} \otimes \bE_{22}+\bE_{22} \otimes \bE_{22} \otimes \bE_{11}\\
&& + g(\theta_{01})g(\theta_{02}) \Big( \bE_{12} \otimes \bE_{12} \otimes \bE_{11} - \bE_{21} \otimes \bE_{21} \otimes \bE_{11} + \bE_{12} \otimes \bE_{21} \otimes \bE_{22}-\bE_{21} \otimes \bE_{12} \otimes \bE_{22}\Big)\nonumber\\
&& + g(\theta_{01})g(\theta_{03}) \Big( \bE_{12} \otimes \bE_{11} \otimes \bE_{12} - \bE_{21} \otimes \bE_{11} \otimes \bE_{21} - \bE_{12} \otimes \bE_{22} \otimes \bE_{21}+\bE_{21} \otimes \bE_{22} \otimes \bE_{12}\Big)\nonumber\\&& + g(\theta_{02})g(\theta_{03}) \Big( \bE_{11} \otimes \bE_{12} \otimes \bE_{12} - \bE_{11} \otimes \bE_{21} \otimes \bE_{21} + \bE_{22} \otimes \bE_{12} \otimes \bE_{21}-\bE_{22} \otimes \bE_{21} \otimes \bE_{12}\Big).\nonumber
\end{eqnarray}
It is easy to see that the fermionic states are all annihilated by $T_3$. On the bosonic subspace, we can again recast the action of $T_3$ in terms of a matrix:
\begin{equation*}
  T_3 = \mathfrak{1} + \, e^{-\theta_0}
  \begin{blockarray}{*{4}{c} l}
    \begin{block}{*{4}{c}>{$\footnotesize}l<{$}}
      |\phi\rangle \otimes |\phi\rangle \otimes |\phi\rangle & |\phi\rangle \otimes |\psi\rangle \otimes |\psi\rangle & |\psi\rangle \otimes |\phi\rangle \otimes |\psi\rangle & |\psi\rangle \otimes |\psi\rangle \otimes |\phi\rangle &\\
    \end{block}
    \begin{block}{[*{4}{c}]>{}}
      0 & t_{23} & t_{13} & t_{12} & |\phi\rangle \otimes |\phi\rangle \otimes |\phi\rangle \\
      -t_{23} & 0 & t_{12} & -t_{13} & |\phi\rangle \otimes |\psi\rangle \otimes |\psi\rangle \\
    -t_{13} & -t_{12} & 0 & t_{23} & |\psi\rangle \otimes |\phi\rangle \otimes |\psi\rangle \\
    -t_{12} & -t_{13} & -t_{23} & 0 & |\psi\rangle \otimes |\psi\rangle \otimes |\phi\rangle \\  
       \end{block}
  \end{blockarray}
\end{equation*}
where also in this case we notice the presence of an antisymmetric matrix carrying the dependence on the quantum spaces, added to an identity piece independent on any variable. 

Once again, it is immediate to see that the bosonic eigenvectors do not depend on $\theta_0$. We shall not need their explicit form here, only the eigenvalues of $T_3$. There are only two of those, each with multiplicity 2:
\begin{eqnarray}
\lambda_\pm = 1 \pm i e^{-\theta_0} \, \Big( e^{\theta_1 + \theta_2} + e^{\theta_1 + \theta_3}+ e^{\theta_2 + \theta_3}\Big)^\frac{1}{2}.
\end{eqnarray}

The auxiliary Bethe equation reads 
\begin{equation}
\tanh \frac{\beta - \theta_1}{2} \, \tanh \frac{\beta - \theta_2}{2} \, \tanh \frac{\beta - \theta_3}{2}= -1,
\end{equation}
with solutions
\begin{eqnarray}
&&e^{\frac{\beta}{2}} = 0 \qquad \mbox{\it i.e.} \, \, \, \, \beta = - \infty, \nonumber \\ &&e^{\frac{\beta}{2}} = \pm e^{i \frac{\pi}{4}} \Big( e^{\theta_1 + \theta_2} + e^{\theta_1 + \theta_3}+ e^{\theta_2 + \theta_3}\Big)^\frac{1}{4} \equiv x_\pm, \nonumber \\ &&e^{\frac{\beta}{2}} = \pm e^{3 i \frac{\pi}{4}} \Big( e^{\theta_1 + \theta_2} + e^{\theta_1 + \theta_3}+ e^{\theta_2 + \theta_3}\Big)^\frac{1}{4} \equiv y_\pm.
\end{eqnarray}
We see that, as in the case of 3 particles, we have again
\begin{equation}
\lambda_- = e^{-\theta_0} \big(e^{\frac{\theta_0}{2}} - x_+\big)\big(e^{\frac{\theta_0}{2}} - x_-\big), \qquad \lambda_+ = e^{-\theta_0} \big(e^{\frac{\theta_0}{2}} - y_+\big)\big(e^{\frac{\theta_0}{2}} - y_-\big).
\end{equation}

\subsubsection{\label{sec:4p}5 particles}

The 4-particle case is clearly not sufficient to suggest the general trend, and we need to work out the case of 5 particles ({\it i.e.} $N=4$) to be in the position to extrapolate. 

The $N=4$ computation is quite legthy, but it is worth illustrating the salient points. One finds that there are two distinct contributions:
\begin{equation}
T_4 = \Omega_1 + \Omega_2.
\end{equation}
The $\Omega_1$ term goes like
\begin{equation}
\Omega_1 = \bE_{11} \otimes \bE_{11} \otimes \bE_{11} \otimes \bE_{11} + \bE_{22} \otimes \bE_{22} \otimes \bE_{22} \otimes \bE_{22} + \Big(\bE_{11} \otimes \bE_{11} \otimes \bE_{22} \otimes \bE_{22} + \mbox{perm.}\Big), 
\end{equation}
where ``perm." denotes all possible arrangements of exactly 2 operators $\bE_{11}$ and 2 operators $\bE_{22}$. The resulting operator $\Omega_1$ clearly annihilates all the fermionic states. 

The term $\Omega_2$ is more complicated to describe. Define the operators  $m_{ij}$, $n_{ij}$, $\gamma_{ij}$ and $\tau_{ij}$, acting at sites $i$ and $j$, by their only non-zero actions
\begin{eqnarray}
&&m_{ij} \, |\phi\rangle_i \otimes |\phi\rangle_j = - g(\theta_{0i})g(\theta_{0j}) \, |\psi\rangle_i \otimes |\psi\rangle_j, \qquad m_{ij} \, |\psi\rangle_i \otimes |\psi\rangle_j = g(\theta_{0i})g(\theta_{0j}) \, |\phi\rangle_i \otimes |\phi\rangle_j,\nonumber\\
&&n_{ij} \, |\phi\rangle_i \otimes |\psi\rangle_j = - g(\theta_{0i})g(\theta_{0j}) \, |\psi\rangle_i \otimes |\phi\rangle_j, \qquad n_{ij} \, |\psi\rangle_i \otimes |\phi\rangle_j = g(\theta_{0i})g(\theta_{0j}) \, |\phi\rangle_i \otimes |\psi\rangle_j,\nonumber\\
&&\rho_{ij} \, |\phi\rangle_i \otimes |\phi\rangle_j = |\phi\rangle_i \otimes |\phi\rangle_j, \qquad \rho_{ij} \, |\psi\rangle_i \otimes |\psi\rangle_j = \, |\psi\rangle_i \otimes |\psi\rangle_j,\nonumber\\
&&\sigma_{ij} \, |\phi\rangle_i \otimes |\psi\rangle_j = |\phi\rangle_i \otimes |\psi\rangle_j, \qquad \sigma_{ij} \, |\psi\rangle_i \otimes |\phi\rangle_j = \, |\psi\rangle_i \otimes |\phi\rangle_j,\nonumber\\
&&\gamma_{ij} \, |\phi\rangle_i \otimes |\phi\rangle_j = |\phi\rangle_i \otimes |\phi\rangle_j, \qquad \gamma_{ij} \, |\psi\rangle_i \otimes |\psi\rangle_j = - \, |\psi\rangle_i \otimes |\psi\rangle_j,\nonumber\\
&&\tau_{ij} \, |\phi\rangle_i \otimes |\psi\rangle_j = |\phi\rangle_i \otimes |\psi\rangle_j, \qquad \tau_{ij} \, |\psi\rangle_i \otimes |\phi\rangle_j = - \, |\psi\rangle_i \otimes |\phi\rangle_j,
\end{eqnarray}
where the index of a state $|v\rangle_i$ indicates that the state is sitting in position $i$. One then finds
\begin{eqnarray}
\Omega_2 &=& m_{12} \, \rho_{34} + n_{12} \, \sigma_{34} + m_{23} \, \rho_{14} + n_{23} \, \sigma_{14} + m_{34} \, \rho_{12} + n_{34} \sigma_{12} + m_{13} \, \gamma_{24} + n_{13} \, \tau_{24} + m_{24} \, \gamma_{13} +  \nonumber\\
&& n_{24} \, \tau_{13} + m_{14} \, \rho_{23} - n_{14} \, \sigma_{23} + m_{12} \, m_{34} + n_{12} \, n_{34},
\end{eqnarray}
which also manifestly annihilates all fermionic states. To re-express this into matrix language, we can introduce a basis of the bosonic subspace of states:
\begin{eqnarray}
&&|1\rangle \equiv |\phi\rangle \otimes |\phi\rangle \otimes |\phi\rangle \otimes |\phi\rangle, \qquad |2\rangle \equiv |\psi\rangle \otimes |\psi\rangle \otimes |\phi\rangle \otimes |\phi\rangle,\nonumber\\
&&|3\rangle \equiv |\psi\rangle \otimes |\phi\rangle \otimes |\psi\rangle \otimes |\phi\rangle, \qquad |4\rangle \equiv |\psi\rangle \otimes |\phi\rangle \otimes |\phi\rangle \otimes |\psi\rangle,\nonumber\\
&&|5\rangle \equiv |\phi\rangle \otimes |\psi\rangle \otimes |\psi\rangle \otimes |\phi\rangle, \qquad |6\rangle \equiv |\phi\rangle \otimes |\psi\rangle \otimes |\phi\rangle \otimes |\psi\rangle,\nonumber\\
&&|7\rangle \equiv |\phi\rangle \otimes |\phi\rangle \otimes |\psi\rangle \otimes |\psi\rangle, \qquad |8\rangle \equiv |\psi\rangle \otimes |\psi\rangle \otimes |\psi\rangle \otimes |\psi\rangle,
\end{eqnarray} 
in terms of which one obtains
\begin{equation}
T_4 = \mathfrak{1} + \, e^{-2 \theta_0 + \frac{1}{2}\sum_{i=1}^4 \theta_i} \, Q + \, e^{-\theta_0} \, G,
\end{equation}
where $Q$ is a symmetric anti-diagonal matrix
\begin{equation*}
  Q =  \begin{blockarray}{*{8}{c} l}
    \begin{block}{*{8}{c}>{$\footnotesize}l<{$}}
      |1\rangle & |2\rangle &|3\rangle &|4\rangle &|5\rangle &|6\rangle &|7\rangle &|8\rangle &\\
    \end{block}
    \begin{block}{[*{8}{c}]>{}}
      0 & 0 & 0 & 0 & 0 & 0 & 0 & 1 & |1\rangle \\
      0 & 0 & 0 & 0 & 0 & 0 & -1 & 0 & |2\rangle \\
      0 & 0 & 0 & 0 & 0 & 1 & 0 & 0 & |3\rangle \\
      0 & 0 & 0 & 0 & -1 & 0 & 0 & 0 & |4\rangle \\
      0 & 0 & 0 & -1 & 0 & 0 & 0 & 0 & |5\rangle \\
      0 & 0 & 1 & 0 & 0 & 0 & 0 & 0 & |6\rangle \\
      0 & -1 & 0 & 0 & 0 & 0 & 0 & 0 & |7\rangle \\
      1 & 0 & 0 & 0 & 0 & 0 & 0 & 0 & |8\rangle \\  
       \end{block}
  \end{blockarray}
\end{equation*}
while $G$ is the antisymmetric matrix
\begin{equation*}
  G =  \begin{blockarray}{*{8}{c} l}
    \begin{block}{*{8}{c}>{$\footnotesize}l<{$}}
      |1\rangle & |2\rangle &|3\rangle &|4\rangle &|5\rangle &|6\rangle &|7\rangle &|8\rangle &\\
    \end{block}
    \begin{block}{[*{8}{c}]>{}}
      0 & t_{12} & t_{13} & t_{14} & t_{23} & t_{24} & t_{34} & 0 & |1\rangle \\
      -t_{12} & 0 & -t_{23} & -t_{24} & t_{13} & t_{14} & 0 & t_{32} & |2\rangle \\
      -t_{13} & t_{23} & 0 & -t_{34} & -t_{12} & 0 & t_{14} & -t_{24} & |3\rangle \\
      -t_{14} & t_{24} & t_{34} & 0 & 0 & -t_{12} & -t_{13} & t_{23} & |4\rangle \\
      -t_{23} & -t_{13} & t_{12} & 0 & 0 & -t_{34} & t_{24} & t_{14} & |5\rangle \\
      -t_{24} & -t_{14} & 0 & t_{12} & t_{34} & 0 & -t_{23} & -t_{13} & |6\rangle \\
      -t_{34} & 0 & -t_{14} & t_{13} & -t_{24} & t_{23} & 0 & t_{12} & |7\rangle \\  
      0 & -t_{34} & t_{24} & -t_{23} & -t_{14} & t_{13} & -t_{12} & 0 & |8\rangle \\
       \end{block}
  \end{blockarray}
\end{equation*}
It is easy to verify that different powers of $e^{-\theta_0}$ are associated to different commuting charges:
\begin{equation}
[Q, G] = 0,
\end{equation}
hence the eigenstates do not depend on $\theta_0$ and are common to $Q$ and $G$. In particular, $Q$ has eigenvalues $\pm 1$, both with multiplicity 4. Once again, we shall not need the specific form of the eigenvectors, and simply report the eigenvalues of $T_4$. They are 4 distinct ones, and read
\begin{eqnarray}
&&\lambda_{1,\pm} = 1 + e^{-2 \theta_0 + \frac{1}{2}\sum_{i=1}^4 \theta_i} \pm i e^{-\theta_0} \, \bigg( \sum_{i<j=1}^4 e^{\theta_i + \theta_j} - 2 e^{\sum_{i=1}^4 \frac{\theta_i}{2}}\bigg)^\frac{1}{2},\nonumber\\
&&\lambda_{2,\pm} = 1 - e^{-2 \theta_0 + \frac{1}{2}\sum_{i=1}^4 \theta_i} \pm i e^{-\theta_0} \, \bigg( \sum_{i<j=1}^4 e^{\theta_i + \theta_j} + 2 e^{\sum_{i=1}^4 \frac{\theta_i}{2}}\bigg)^\frac{1}{2},
\end{eqnarray}
each with multiplicity 2.

The auxiliary Bethe equation reads 
\begin{equation}
\prod_{i=1}^4 \tanh \frac{\beta - \theta_i}{2} = -1,
\end{equation}
with solutions
\begin{eqnarray}
&&e^{\frac{\beta}{2}} = \pm \frac{e^{i \frac{\pi}{4}}}{2^\frac{1}{4}} \bigg( Z^\frac{1}{2} + \sum_{i<j=1}^4 e^{\theta_i + \theta_j} \bigg)^\frac{1}{4} \equiv x_\pm, \qquad e^{\frac{\beta}{2}} = \pm \frac{e^{3 i \frac{\pi}{4}}}{2^\frac{1}{4}} \bigg( Z^\frac{1}{2} + \sum_{i<j=1}^4 e^{\theta_i + \theta_j} \bigg)^\frac{1}{4} \equiv y_\pm, \\ 
&&e^{\frac{\beta}{2}} = \pm \frac{e^{i \frac{\pi}{4}}}{2^\frac{1}{4}} \bigg( - Z^\frac{1}{2} + \sum_{i<j=1}^4 e^{\theta_i + \theta_j} \bigg)^\frac{1}{4} \equiv \tilde{x}_\pm, \qquad e^{\frac{\beta}{2}} = \pm \frac{e^{3 i \frac{\pi}{4}}}{2^\frac{1}{4}} \bigg( - Z^\frac{1}{2} + \sum_{i<j=1}^4 e^{\theta_i + \theta_j} \bigg)^\frac{1}{4} \equiv \tilde{y}_\pm,\nonumber
\end{eqnarray}
where
\begin{equation}
Z = \bigg( \sum_{i<j=1}^4 e^{\theta_i + \theta_j}\bigg)^2 - 4 \, e^{\sum_{i=1}^4 \theta_i}.
\end{equation}
We see that now
\begin{eqnarray}
\label{pro}
&& \lambda_{2,-} = e^{-2\theta_0} \big(e^{\frac{\theta_0}{2}} - x_+\big)\big(e^{\frac{\theta_0}{2}} - x_-\big)\big(e^{\frac{\theta_0}{2}} - \tilde{x}_+\big)\big(e^{\frac{\theta_0}{2}} - \tilde{x}_-\big), \nonumber \\
&& \lambda_{1,-} = e^{-2\theta_0} \big(e^{\frac{\theta_0}{2}} - x_+\big)\big(e^{\frac{\theta_0}{2}} - x_-\big)\big(e^{\frac{\theta_0}{2}} - \tilde{y}_+\big)\big(e^{\frac{\theta_0}{2}} - \tilde{y}_-\big), \nonumber \\
&& \lambda_{1,+} = e^{-2\theta_0} \big(e^{\frac{\theta_0}{2}} - y_+\big)\big(e^{\frac{\theta_0}{2}} - y_-\big)\big(e^{\frac{\theta_0}{2}} - \tilde{x}_+\big)\big(e^{\frac{\theta_0}{2}} - \tilde{x}_-\big),\nonumber\\
&& \lambda_{2,+} = e^{-2\theta_0} \big(e^{\frac{\theta_0}{2}} - y_+\big)\big(e^{\frac{\theta_0}{2}} - y_-\big)\big(e^{\frac{\theta_0}{2}} - \tilde{y}_+\big)\big(e^{\frac{\theta_0}{2}} - \tilde{y}_-\big),
\end{eqnarray}
where we have used the double-radical formula
\begin{equation}
\sqrt{A + \sqrt{B}} \pm \sqrt{A - \sqrt{B}} = \sqrt{2} \, \sqrt{A \pm \sqrt{A^2 - B}}
\end{equation}
in performing the products in (\ref{pro}).

\subsubsection{\label{sec:Conj}Conjecture for generic $N$}

We are now ready to attempt a conjecture for generic $N$. On the one hand, controlling more low-$N$ cases would certainly provide us with more confidence in our intuition, on the other hand already $N=5$ looks computationally rather intimidating. Nevertheless, a trend seems to emerge with sufficient clarity for us to dare extrapolating, although either more checks or an analytic proof ({\it e.g.} using induction) would obviously be desirable.  

We also notice that, in all cases $N=1,2,3,4$ which we have explored, we have always been able to confirm the {\it inversion relation} found in \cite{Fontanella:2017rvu}:
\begin{equation}
\lambda(\theta_0) \, \lambda(\theta_0 + i \pi) = \frac{1}{4} \, \big(1+F\big) \Bigg[\prod_{i=1}^N \big(1 + e^{-\theta_0 + \theta_i}\big) + \prod_{i=1}^N \big(1 - e^{-\theta_0 + \theta_i}\big)\Bigg],
\end{equation}
where we have taken into account the factor $\frac{1}{2}$ in $T_N$, and of course $F=1$ for bosonic states, $F=-1$ for fermionic ones. In particular, we explicitly confirmed that all fermionic states are annihilated by the transfer matrix built using our procedure. Moreover, all the eigenvalues we have found are indeed meromorphic and $2 \pi i $-periodic in $\theta_0$.

We conjecture the bosonic eigenvalues (with reinstated dressing factors $\Omega$) at generic $N$ to simply go like
\begin{eqnarray}
\label{ei}
\lambda = \prod_{i=1}^N \Omega(\theta_0 - \theta_i) \, \prod_{\mbox{choices}}^{[\frac{N}{2}]} \Big(1 - e^{- \theta_0} e^{\beta_m}\Big),
\end{eqnarray}
with $[n]$ denoting the integer part of $n$, {\it e.g.} $[\frac{3}{2}] = 1$. Here ``choices" means all possibile choices of $[\frac{N}{2}]$ solutions $\beta_m$ of the auxiliary Bethe equations
\begin{equation}
\prod_{i=1}^N \tanh \frac{\beta - \theta_i}{2} = -1,
\end{equation}
with imaginary part $\mathfrak{Im}(\beta) \in (0,2 \pi)$ when all the $\theta_i$ are set to 0, and with {\it distinct real parts} when all the $\theta_i$ are set to 0. In an alternative terminology this could be phrased as being a product over distinct {\it centres} of the auxiliary Bethe roots. 

In fact, the auxiliary Bethe equations admit two classes of solutions, singled out by having either $\mathfrak{Im}(\beta) = \frac{\pi}{2}$ or $\mathfrak{Im}(\beta) = \frac{3 \pi}{2}$ when all the $\theta_i$ are set to 0. This fits with the fact the auxiliary Bethe equations are invariant (each sides becoming its own reciprocal) under $\beta \to \beta + i \pi$.

Notice that we have chosen to pairwise combine terms corresponding to auxiliary Bethe roots $x_+$ and $x_-$, $y_+$ and $y_-$, etc., which explains the $[\frac{N}{2}]$ counting, and the presence of $\theta_0$ and $\beta$ instead of $\frac{\theta_0}{2}$ and $\frac{\beta}{2}$.

\subsection{\label{sec:MainNonRel}Non-relativistic}

After having performed the analysis for the relativistic case, it is simple to extend our results to the non-relativistic situation. As explained in section \ref{sec:AuxNonRel}, the auxiliary Bethe equations are rather easily found by the simple replacement (\ref{repl}), which works for the eigenvalues of the transfer matrix as well. One in fact has now (\ref{S}) replaced by $S(p_1,p_2)$, as a consequence of\footnote{It is also easy to check that $g(p_1,p_2) \to g(\theta_1-\theta_2)$ if $p_i  = \epsilon \, e^{\theta_i}$, $\epsilon \to 0$ (relativistic limit).} 
\begin{equation}
g(\theta_{12}) \qquad \mbox{\underline{replaced by}} \qquad g(p_1,p_2) = \sqrt{\frac{\tan \frac{p_2}{4}}{\tan \frac{p_1}{4}}},
\end{equation}
and in (\ref{T}) all the S-matrices
\begin{equation}
S_{0i}(\theta_{0i}) \qquad \mbox{\underline{replaced by}} \qquad S_{0i}(p_0,p_i)
\end{equation}
accordingly. The structure is largely the same, hence the algebraic manipulations can be borrowed from the relativistic analysis. The final step, introducing the solutions of the auxiliary Bethe equations and rewriting the eigenvalues in terms of such solutions, is only marginally trickier. We very briefly report the analysis here below for up to 4 particles, when it will become clear how to generalise the formulas.

Once again, the reader interested in the final formulas can skip to section \ref{sec:NRConj} in the first instance.

\subsubsection{\label{sec:NR1p}2 particle}
The eigenvalues is equal to $1$ as in the relativistic 2-particle case. The auxiliary Bethe equation
\begin{equation}
\frac{\sin \frac{q + p_1}{4}}{\sin \frac{q - p_1}{4}} = - 1
\end{equation}
has solutions
\begin{equation}
e^{i \frac{q}{4}} = \pm 1.
\end{equation}

\subsubsection{\label{sec:NR2p}3 particles}

The eigenvalues of the transfer matrix for 3 particles read
\begin{equation}
\lambda_\pm = \Big(1 \pm i g(p_0,p_1) g(p_0,p_2)\Big).
\end{equation}
The auxiliary Bethe equation
\begin{equation}
\frac{\sin \frac{q + p_1}{4}}{\sin \frac{q - p_1}{4}}\, \, \frac{\sin \frac{q + p_2}{4}}{\sin \frac{q - p_2}{4}} = - 1
\end{equation}
has solutions
\begin{equation}
\tan \frac{q}{4} = i \sqrt{\tan \frac{p1}{4} \, \tan \frac{p2}{4}} \equiv x, \qquad \tan \frac{q}{4} = -i \sqrt{\tan \frac{p1}{4} \, \tan \frac{p2}{4}} \equiv y.
\end{equation}
One can verify that
\begin{equation}
\lambda_+ = 1 - y \, \cot \frac{p_0}{4}, \qquad \lambda_- = 1 - x \, \cot \frac{p_0}{4}.
\end{equation}

\subsubsection{\label{sec:NR3p}4 particles}

The eigenvalues of the transfer matrix for 4 particles read
\begin{equation}
\lambda_\pm = 1 \pm i \sqrt{g(p_0,p_1)^2 g(p_0,p_2)^2 + g(p_0,p_1)^2 g(p_0,p_3)^2 + g(p_0,p_2)^2 g(p_0,p_2)^2}.
\end{equation}
The auxiliary Bethe equation
\begin{equation}
\frac{\sin \frac{q + p_1}{4}}{\sin \frac{q - p_1}{4}}\, \, \frac{\sin \frac{q + p_2}{4}}{\sin \frac{q - p_2}{4}}\, \, \frac{\sin \frac{q + p_3}{4}}{\sin \frac{q - p_3}{4}} = - 1
\end{equation}
has solutions
\begin{equation}
e^{i \frac{q}{4}} = \pm 1, \qquad \tan \frac{q}{4} = x \equiv - i \frac{b - \frac{1}{b}}{b + \frac{1}{b}}, \qquad \tan \frac{q}{4} = y \equiv - i \frac{a - \frac{1}{a}}{a + \frac{1}{a}},
\end{equation}
where
\begin{eqnarray}
&&b = \nonumber\\
&&\frac{\sqrt{- 1 + \sum_{i=1}^3 \zeta_i \, + \sum_{i<j} \zeta_i \zeta_j - \zeta_1 \zeta_2 \zeta_3 + \sqrt{\Big[- 1 + \sum_{i=1}^3 \zeta_i \, + \sum_{i<j} \zeta_i \zeta_j - \zeta_1 \zeta_2 \zeta_3 \Big]^2 - 4 \Big[1 + \zeta_1 \zeta_2 \zeta_3 \Big]^2}}}{\sqrt{2 \big( 1 + \zeta_1 \zeta_2 \zeta_3\big)}},\nonumber\\
&&a = \nonumber\\
&&\frac{\sqrt{- 1 + \sum_{i=1}^3 \zeta_i \, + \sum_{i<j} \zeta_i \zeta_j - \zeta_1 \zeta_2 \zeta_3 - \sqrt{\Big[- 1 + \sum_{i=1}^3 \zeta_i \, + \sum_{i<j} \zeta_i \zeta_j - \zeta_1 \zeta_2 \zeta_3 \Big]^2 - 4 \Big[1 + \zeta_1 \zeta_2 \zeta_3 \Big]^2}}}{\sqrt{2 \big( 1 + \zeta_1 \zeta_2 \zeta_3\big)}},\nonumber\\
\end{eqnarray}
and
\begin{equation}
\zeta_i = e^{i \frac{p_i}{4}}.
\end{equation}
One can check that
\begin{equation}
\lambda_+ = 1 - y \, \cot \frac{p_0}{4}, \qquad \lambda_- = 1 - x \, \cot \frac{p_0}{4}.
\end{equation}

\subsubsection{\label{sec:NRConj}Conjecture for generic $N$}

The case of 5 particles is prohibitively complicated to deal with, nevertheless the experience with the relativistic case appears to be sufficient to suggest a generalisation of the previous results. In fact, one can see that a clear correspondence persists between the type of solutions of the auxiliary Bethe equations in the two cases.

We therefore extend our conjecture to the non-relativistic case by writing
\begin{equation}
\lambda = \prod_{i=1}^N \Omega(p_0,p_i) \, \prod_{\mbox{choices}}^{[\frac{N}{2}]} \Big(1 - \cot \frac{p_0}{4} \, \tan \frac{q_m}{4}\Big),
\end{equation}
with $[n]$ denoting the integer part of $n$. Here ``choices" means all possibile choices of $[\frac{N}{2}]$ solutions $q_m$ of the auxiliary Bethe equations
\begin{equation}
\prod_{i=1}^N \frac{\sin \frac{q + p_i}{4}}{\sin \frac{q - p_i}{4}} = -1
\end{equation}
which respect the pattern of section \ref{sec:Conj} {\it in their relativistic limit}. 

We have also inserted the non-relativistic dressing factors $\Omega(p_0,p_i)$, which we will discuss in a later section.

\section{\label{sec:Asym}Massless asymptotic Bethe ansatz}

We can finally write down the complete asymptotic Bethe ansatz, which is now a simple task if we rely on the arguments contained in section \ref{sec:Main}. In fact, if one considers $N+1$ particles on a circle of length $L$ with periodic boundary conditions, interacting with one another via an integrable scattering matrix, one is brought to impose the following constituent equation - see \cite{Ahn:1993qa}, eq.s (3.8) and (3.9):
\begin{eqnarray}
\label{eige}
e^{i p_0 L} \, M(p_0 | p_1,...,p_N) |\psi\rangle = |\psi \rangle, 
\end{eqnarray} 
where $p_i$ is the momentum of the $i$-th particle on the circle
\begin{eqnarray}
p_i = e^{\theta_i}, \qquad i=0,...,N,
\end{eqnarray} 
while
\begin{eqnarray}
\label{tracea}
T(p_0|p_1,...,p_N) = \mbox{tr}_0 M(p_0|p_1,...,p_N)
\end{eqnarray} 
is the transfer matrix. The physical situation is that of revolving particle $0$ around the circle of length $L$ while scattering all the other ones in sequence, which amounts to the identity acting on an eigenstate $|\psi \rangle$ of the monodromy matrix $M$. 

We also have to recall that we divided by $2$ in the definition of the transfer matrix, and we have now to reinstate that factor when writing the Bethe equations. This gets rid of the factor $2$ which would appear from the trace, corresponding to $N_c$ in eq. (3.9) of \cite{Ahn:1993qa}.

\subsection{\label{sec:AsymRel}Relativistic}

We therefore write the core equations of the Bethe ansatz, with the same notations of section \ref{sec:Conj},
\begin{eqnarray}
\label{BetheEq}
&&e^{i e^{\theta_0} L} \, \prod_{b=1}^N \Omega(\theta_0 - \theta_b) \, \prod_{\mbox{choices}}^{[\frac{N}{2}]} 
\Big(1 - e^{- \theta_0} e^{\beta_m}\Big)=1,\\
&&\label{Bethe2}\prod_{b =1}^N \tanh \frac{\beta_m - \theta_b}{2} = -1.
\end{eqnarray}

In \cite{Fontanella:2017rvu}, a minimal solution for the dressing factor was found:
\begin{eqnarray}
\label{omega}
\Omega(\theta) = \frac{e^{\frac{\gamma}{2}- \frac{\pi i }{8}+\frac{\theta}{4}}}{\sqrt{2 \pi}} \prod_{j=1}^\infty e^{-\frac{1}{2 j}} \, j \, \frac{\Gamma\Big(j-\frac{1}{2}+\frac{\theta}{2\pi i}\Big)\Gamma\Big(j-\frac{\theta}{2\pi i}\Big)}{\Gamma\Big(j+\frac{1}{2}-\frac{\theta}{2\pi i}\Big)\Gamma\Big(j+\frac{\theta}{2\pi i}\Big)},
\end{eqnarray}
which satisfies the cross-unitarity equation
\begin{equation}
\Omega (\theta) \Omega(\theta + i \pi) = \frac{e^{\frac{\theta}{2}}}{2 \cosh \frac{\theta}{2}},
\end{equation}
where $\gamma$ is the Euler-Mascheroni constant. The attribute of {\it minimal} for $\Omega(\theta)$ stems from the fact that it has no poles in the physical strip $\theta \in (0,\pi)$, as massless particles cannot form bound states. Hence, no CDDs are necessary.

We remind the reader that the above is for right-moving particles. For left movers, the S-matrix is simply the transpose of the one for right movers, while the mixed S-matrix is trivial (right and left movers decouple in the relativistic limit). The transpose is in the sense of $4 \times 4$ matrices, but can equivalently be seen as $\tr \otimes \tr$ on the individual two-dimensional representations. Hence, the transfer matrix has exactly the same eigenvalues for the left movers as for the right movers. The same holds for the crossing equation for left movers, which coincides with the one for right movers. Therefore, the same dressing factor can be chosen. All this results in the exact same Bethe ansatz for both types of excitations. 

\subsection{\label{sec:AsymNonRel}Non-relativistic}

The non-relativistic case is given accordingly by
\begin{eqnarray}
\label{NRBAE}
&&e^{i p_0 L} \prod_{b=1}^N \Omega(p_0,p_b) \, \prod_{\mbox{choices}}^{[\frac{N}{2}]} \Big(1 - \cot \frac{p_0}{4} \, \, \tan \frac{q_m}{4}\Big) = 1,\\
&&\prod_{b=1}^N \frac{\sin \frac{q_m + p_b}{4}}{\sin \frac{q_m - p_b}{4}} = -1.\nonumber
\end{eqnarray}

Let us study the dressing factor $\Omega(p_0,p_i)$ appearing in (\ref{NRBAE}). The cross-unitarity equation for the S-matrix implies
\begin{equation}
\label{used}
\Omega(p_a,p_b) \Omega(\bar{p}_a,p_b) = \frac{\sin \frac{p_a}{4} \, \, \cos \frac{p_b}{4}}{\sin \frac{p_a+p_b}{4}},
\end{equation}
where 
\begin{equation}
\label{ineq}
\bar{p}_a \equiv p^{(1)}_a
\end{equation}
is the {\it crossed momentum}. More precisely, in (\ref{ineq}) we have introduced the shorthand notation $p^{(n)}_a$ to denote in fact the operation of continuing the momentum $p_a$ to the $n$-th sheet in the complex plane or, equivalently, its uniformising rapidity to the $n$-th region in the complex plane.

We now recall that the crossing equation for the $AdS_3$ massless dressing factor can be written in terms of a function $\sigma(p_a,p_b)$ satisfying \cite{Borsato:2016xns}:
\begin{equation}
\label{usat}
\sigma(p_a,p_b) \, \sigma(\bar{p}_a,p_b) = \frac{\sin \frac{p_a-p_b}{4}}{\sin \frac{p_a+p_b}{4}}.
\end{equation}
This means that we can formally require 
\begin{equation}
\label{folding}
\sigma(p_a,p_b) = \frac{\Omega(p_a,p_b)}{\Omega(\bar{p}_a,p_b)}, 
\end{equation}
since, in this way, we have
\begin{equation}
\sigma(p_a,p_b) \, \sigma (\bar{p}_a,p_b) = \frac{\Omega(p_a,p_b)}{\Omega(\bar{\bar{p}}_a,p_b)} = \frac{\sin \frac{p_a-p_b}{4}}{\sin \frac{p_a+p_b}{4}},
\end{equation} 
where we have combined (\ref{used}) together with its crossed version\footnote{As opposed to $\Omega$, which is a multi-valued function of the momenta (in other words, it is a single-valued function of the uniformising rapidities), the r.h.s.s of (\ref{used}) and (\ref{usat}) truly are functions of the momenta themselves. This means that, under crossing, we can simply replace $\bar{p}_a$ with $- p_a$ in the arguments of the trigonometric functions appearing there.}, and we have again used 
\begin{equation}
\bar{\bar{p}}_a \equiv p^{(2)}_a
\end{equation}
as a shorthand notation for the crossing transformation having been performed twice, which moves the uniformising rapidity-variable one region further. This will be consistent with the respective crossing equations.

We consider formula (\ref{folding}) as the non-relativistic corrispective of Melzer's {\it folding} \cite{Melzer}, which connects the dressing factor of ${\cal{N}}=2$ supersymmetric S-matrices with the one of ${\cal{N}}=1$ S-matrices. In fact, the $AdS_3$ scattering problem is related to ${\cal{N}}=2$ supersymmetry \cite{Diego}, while we know that the $AdS_2$ one is associated to ${\cal{N}}=1$ supersymmetry. A similar folding relation is then expected to extend to the thermodynamic Bethe ansatz as well.

We can then formally invert the relation (\ref{folding}), to find 
\begin{equation}
\label{expro}
\Omega(p_a,p_b) = \prod_{n=0}^\infty \sigma\Big(p_a^{(n)},p_b\Big).
\end{equation} 

The details of the analytic continuation implementing the crossing map, as well as the explicit form of $\sigma (p_a,p_b)$, can be found in \cite{Borsato:2016xns}. Here, we simply remark that the expression (\ref{expro}) is unfortunately only formal at the moment, as it is quite hard to ascertain the convergence of the infinite product - which might require to be regularised\footnote{For instance, rearranging the product pairwise in (\ref{expro}) gives
\begin{equation}
\lim_{N \to \infty} \bigg[\frac{\sin \frac{p_a+p_b}{4}}{\sin \frac{p_a-p_b}{4}}\bigg]^N,
\end{equation} 
which is either $0$, or $1$, or it diverges. The reason behind this is that inverting (\ref{folding}) is not a unique operation, the most general solution being 
\begin{equation}
\label{exproo}
\Omega(p_a,p_b) = C \, \prod_{n=0}^\infty w(n) \, \, \sigma\Big(p_a^{(n)},p_b\Big).
\end{equation}
where the constant $C$ and the function $w(n)$ do not depend on $p_a$ or $p_b$. More information is needed to fix $C$ and $w(n)$. They play the same role as the factor $e^{-\frac{1}{2 j}} \, j$ inside the product in (\ref{omega}).}. We plan to come back to this issue in the future.

Let us also point out that the above is for right-right scattering ({\it i.e.}, only right moving particles in the game). The left-left scattering produces exactly the same formulas. In fact, it can easily be seen that the left-left transfer matrix is the traspose of the right-right one, hence the eigenvalues when only left movers particles are considered will be the same.

However, as opposed to the relativistic case, now we have a non trivial mixed right-left scattering. This means that the right and left sector do not decouple, and one can have mixed configurations. The various possibilities should then be encompassed by a set of variables which can simultaneously incorporate right and left movers. These are the $x$ variables in \cite{Borsato:2016xns}. We expect the full Bethe ansatz for right {\it and} left movers to be obtained from the formula
\begin{eqnarray}
\label{fullNRBAE}
&&e^{i p_0 L} \prod_{b=1}^N \Omega(p_0,p_b) \, \prod_{\mbox{choices}}^{[\frac{N}{2}]} \bigg(1 - \frac{x_0 + 1}{x_0 - 1} \, \, \frac{z_m - 1}{z_m + 1}\bigg) = 1,\nonumber\\
&&\prod_{b=1}^N \frac{1- x_b z_m}{x_b - z_m} = -1,
\end{eqnarray}
by specifying right- or left-moving kinematics for each variable:
\begin{equation}
\label{cove}
x_a = \pm e^{i \frac{p_a}{2}}, \qquad a=0,...,N, \qquad \qquad z_m = \pm e^{i \frac{q_m}{2}},
\end{equation}
where $+$ is for right, $-$ for left movers\footnote{\label{footno}We remind the reader that we have chosen the conventions (\ref{cove}), and not those of having positive momenta comprised between $\mathfrak{Re} p \in (0,2 \pi)$, and having right movers with $\mathfrak{Re} p \in (0, \pi)$ and left movers with $\mathfrak{Re} p \in (\pi ,2 \pi)$, as instead \cite{Borsato:2016xns} did.}. We also consider formula (\ref{folding}) to be applicable with each respective choices of right and left movers, since \cite{Borsato:2016xns} does indeed give an expression for $\sigma$ which is valid for all choices of kinematics (with the caveat of footnote \ref{footno}). 

\subsection{\label{sec:Full}Full Dynkin diagram}
We finally recall that the entire spectrum of $AdS_2 \times S^2 \times T^6$ superstring is governed by two copies of the centrally-extended $\alg{psu}(1|1)$ scattering problem we have just described. Given the factorised structure of the complete S-matrix, the eigenstates of the transfer matrix are simply the tensor product of two copies of the ones we have been discussing, and the eigenvalues are multiplied. We therefore recast our conjecture into the familiar structure of the two-winged $\alg{psu}(1,1|2)$ Dynkin diagram, with a shared momentum-carrying central node (conventionally labelled as node $2$) and two auxiliary ones (labelled $1$ and $3$, respectively):

\begin{itemize}

\item \underline{relativistic}
\begin{eqnarray}
\label{BetheEqo}
&&\label{Bethe2o}\prod_{b =1}^N \tanh \frac{\beta_{1,m} - \theta_b}{2} = -1, \nonumber\\
&&e^{i e^{\theta_0} L} \, \prod_{b=1}^N \Omega^2(\theta_0 - \theta_b) \, \prod_{\mbox{choices}}^{[\frac{N}{2}]} 
\Big(1 - e^{- \theta_0} e^{\beta_{1,m}}\Big)\prod_{\mbox{choices}}^{[\frac{N}{2}]} 
\Big(1 - e^{- \theta_0} e^{\beta_{3,m}}\Big)=1,\nonumber\\
&&\label{Bethe2oo}\prod_{b =1}^N \tanh \frac{\beta_{3,m} - \theta_b}{2} = -1;
\end{eqnarray}
\medskip
\item \underline{non-relativistic}
\begin{eqnarray}
\label{fullNRBAEa}
&&\prod_{b=1}^N \frac{1- x_b z_{1,m}}{x_b - z_{1,m}} = -1, \nonumber\\
&&e^{i p_0 L} \prod_{b=1}^N \Omega^2(p_0,p_b) \, \prod_{\mbox{choices}}^{[\frac{N}{2}]} \bigg(1 - \frac{x_0 + 1}{x_0 - 1} \, \, \frac{z_{1,m} - 1}{z_{1,m} + 1}\bigg)\prod_{\mbox{choices}}^{[\frac{N}{2}]} \bigg(1 - \frac{x_a + 1}{x_a - 1} \, \, \frac{z_{3,m} - 1}{z_{3,m} + 1}\bigg) = 1,\nonumber\\
&&\prod_{b=1}^N \frac{1- x_b z_{3,m}}{x_b - z_{3,m}} = -1.
\end{eqnarray}
\end{itemize}
\bigskip

\begin{figure}
  \centering

    \begin{tikzpicture}
      [
      thick,
      node/.style={shape=circle,draw,thick,inner sep=0pt,minimum size=5mm}
      ]

      \useasboundingbox (-1.5cm,-1cm) rectangle (1.5cm,1cm);

      \node (v1) at (-1.1cm, 0cm) [node] {};
      \node (v2) at (  0.0cm, 0cm) [node] {};
      \node (v3) at (  1.1cm, 0cm) [node] {};

      \draw (v1) -- (v2);
      \draw (v2) -- (v3);

\end{tikzpicture}
\caption{Sketch of Dynkin diagram describing the complete massless Bethe ansatz. The grading of the nodes is not immediately manifest from the form of the Bethe equations.}
  \label{fig:dynkin}
\end{figure}
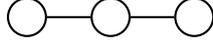

In fig. \ref{fig:dynkin}, we sketch a Dynkin diagram which could describe the above Bethe ansatz. From our procedure, the grading assignement of the nodes is not immediately transparent.
\subsection{\label{bandf}Bosons and fermions}

We end this part with an important remark. It is slightly unsettling that the results we have described up to this point are, strictly speaking, only sensitive to eigenstates of the transfer matrix which are globally bosonic. This however has an explanation. So far we have been concentrating, for definiteness, on what was named solution 3 in \cite{Fontanella:2017rvu}. There is a second S-matrix that appears in the massless limit, according to a choice one has to perform in the reduction from the massive case. As explained in \cite{Fontanella:2017rvu}, this has a perfect correspondence with the two solutions which one finds in ${\cal{N}}=1$ supersymmetric theories \cite{Fendley:1990cy}. In principle, we should now repeat the whole analysis for the second solution (dubbed {\it solution 5} in \cite{Fontanella:2017rvu}). However, it is easy to see that the relativistic S-matrices to be used in the Bethe ansatz can be mapped into one another by the simple exchange
\begin{equation}
\label{mappa}
|\phi\rangle \leftrightarrow |\psi\rangle, \qquad \theta \to - \theta, 
\end{equation} 
where $\theta = \theta_1 - \theta_2$. Therefore, the transfer-matrix eigensystem for solution 5 can simply be obtained by interchanging bosons with fermions, and changing the sign of all the rapidities $\theta_a$, $a=0,1,...,N$ (and \underline{only} of those, not of the $\beta$'s). This generates a very similar Bethe ansatz for fermionic states\footnote{In reality, only for fermionic states with odd $N$, since the map (\ref{mappa}) just shuffles bosonic states with even $N$ among themselves.}. 

The same applies to the non-relativistic case. Here, the map between the S-matrices involves
\begin{equation}
|\phi\rangle \leftrightarrow |\psi\rangle, \qquad \frac{\tan \frac{p_1}{4}}{\tan \frac{p_2}{4}} \to \frac{\tan \frac{p_2}{4}}{\tan \frac{p_1}{4}}.
\end{equation}
Hence, the transfer-matrix eigensystem for the second choice of non-relativistic S-matrix can simply be obtained by interchanging bosons with fermions, and transforming all the variable $x_a \to - \frac{1}{x_a}$, $a=0,1,...,N$ - and \underline{only} those, not the $z$'s.  

\section{\label{sec:Yang}Yangian symmetry and scale invariance}

In this section, we show that the massless relativistic scattering problem enjoys a very peculiar type of Yangian symmetry. This is inherited from the non-relativistic problem \cite{Hoare:2014kma,Hoare:2014kmaa}, however it acquires in this limit a very special property. The Yangian representation in Drinfeld's first realisation \cite{Drinfeld:1985rx} is in fact such that the spectral parameter is exactly equal to zero. In the second realisation \cite{Dsec}, this translates in the corresponding shifted spectral parameter coinciding with the particle's energy. 

This occurrence is quite remarkable. In the last section, we speculate about its possible connection with scale invariance.

\subsection{\label{sec:Yangian}Yangian symmetry}

We first proceed to study the Yangian symmetry of the problem, using the two presentations given by Drinfeld. For the remainder of the paper, we switch to a convention where fermionic signs are re-instated.

\subsubsection{\label{sec:DII}Yangian in Drinfeld's second realisation}

We will use the same Hopf-algebra conventions as in \cite{Fontanella:2017rvu}, which we refer to for the relevant definitions.

The Yangian we consider is generated by $e_n$, $f_n$, $h_n$, with $n\geq 0$ integer, satisfying the following non-zero (anti-)commutation relations in Drinfeld's second realisation,
\begin{equation}
  \label{eq:Lie}
  \begin{gathered}
        \acomm{e_m}{f_n} = 2 h_{m+n}, \qquad \acomm{e_m}{e_n} = 2 P_{m+n}, \qquad 
    \acomm{f_m}{f_n} = 2 P^\dagger_{m+n}.
  \end{gathered}
\end{equation}

The assignement of level-0 generators w.r.t. \cite{Fontanella:2017rvu} works as follows:
\begin{eqnarray}
e_0 = \mathfrak{Q}, \qquad f_0 = \mathfrak{S}, \qquad h_0 = \mathfrak{C}, \qquad P = \mathfrak{P}, \qquad P^\dagger = \mathfrak{K}. 
\end{eqnarray} 
The representation on the fundamental multiplet of one boson and one fermion $\{|\phi\rangle, |\psi\rangle\}$ can also be found in \cite{Fontanella:2017rvu}, as well as the explicit formulas for the R-matrices. The dependence on the single-particle rapidity $\theta$ is encoded in the supercharges $\mathfrak{Q}$ and $\mathfrak{S}$ being both proportional to $e^{\frac{\theta}{2}}$.
 
One can show that the following level-1 Yangian coproducts satisfy the algebra defining relations:
\begin{equation}
\label{from}
  \begin{aligned}
    \Delta(e_1) &= e_1 \otimes \matId + \matId \otimes e_1 +4 \, e_0 \otimes h_0, \\
    \Delta(f_1) &= f_1 \otimes \matId + \matId \otimes f_1 +4 \, h_0 \otimes f_0.
  \end{aligned}
\end{equation}
The coproducts on the central elements $h_n, P_n, P^\dagger_n$ can be obtained from (\ref{from}) by the homomorphism property $\{\Delta(a),\Delta(b)\}=\Delta\big(\{a,b\}\big)$, for $a,b$ any two Yangian fermionic generators.

In order for the coproducts (\ref{from}) to by symmetries of the R-matrix, we need to take the so-called {\it evaluation representation}:
\begin{eqnarray}
J_n = u^n \, J_0,
\end{eqnarray}
where $J$ symbolically denotes any generator. The complex variable $u$ is called {\it evaluation} or {\it spectral parameter}. We discover that symmetry of the R-matrix is achieved only if we choose
\begin{equation}
\label{upper}
u = E = \pm e^\theta,
\end{equation} 
namely, if the spectral parameter equals to eigenvalues $E$ of the energy generator $h_0$ (the upper sign in (\ref{upper}) is for right movers, the lower one for left movers, and we have set the speed of light $c=1$ throughout this paper). This makes the complete Yangian representation in Drinfeld's second realisation very easy to write down:
\begin{equation}
h_n = h^{n+1}, \qquad e_n = h_n e_0, \qquad f_n = h_n f_0.
\end{equation}

The crossing properties of the Yangian generators can easily be obtained from the knowledge of the coproduct, and from the behaviour under crossing of the corresponding level-0 generators. By applying the Hopf-algebra axioms one finds, for the antipode $\mathscr{S}$,
\begin{equation}
\label{plu}
  \mathscr{S} (e_1) = - e_1 +4 \,  e_0 \, h_0, \qquad 
  \mathscr{S} (f_1) = - f_1 +4 \,  f_0 \, h_0.
\end{equation}
Quite interestingly, in evaluation representation we see that
\begin{equation}
\label{ggin}
\mathscr{S} (e_1) = e_1, \qquad \mathscr{S} (f_1) = f_1. 
\end{equation}
This is compatible with the general formula
\begin{equation}
  \label{eq:feed}
  \mathscr{S} \big(J\big) = \mathscr{C}^{-1} \bar{J}^{str} \mathscr{C},
\end{equation}
where $\mathscr{C}$ is the charge-conjugation matrix \cite{Fontanella:2017rvu}, $str$ denotes supertransposition, $J$ is any generator of the Yangian, and the barred representation is given by the antiparticle map 
\begin{equation}
\label{shif}
\theta \to \theta + i \pi.
\end{equation}
The proof of compatibility is easily achieved by noticing that (\ref{eq:feed}) holds at level 0, namely
\begin{equation}
  \label{eq:feedo}
  \mathscr{S} \big(e_0\big) = - e_0 =  \mathscr{C}^{-1} \bar{e_0}^{str} \mathscr{C}.
\end{equation}
Therefore, plugging (\ref{ggin}) into (\ref{eq:feed}), reduces the latter to 
\begin{equation}
u = - \bar{u},
\end{equation} 
which is satisfied by (\ref{upper}) and (\ref{shif}).

\subsubsection{\label{sec:DI}Yangian in Drinfeld's first realisation}

In this case, it is relatively straightforward to change basis of generators to reach Drinfeld's first realisation. It is sufficient to map the level-0 generators into themselves, and to shift the level-1 generators by a bilinear in the level-0 ones, such that the new coproduct-tail is more symmetric. 

By using the homomorphism property, we find that such a map must be given by
\begin{equation}
\label{shift}
\hat{e} = e_1 - 2 e_0 h_0, \qquad \hat{f} = f_1 - 2 f_0 h_0,
\end{equation}
which produces
\begin{equation}
\label{from2}
  \begin{aligned}
    \Delta(\hat{e}) &= \hat{e} \otimes \matId + \matId \otimes \hat{e} +2 \, e_0 \otimes h_0 - 2 \, h_0 \otimes e_0, \\
    \Delta\big(\hat{f} \,\big) &= \hat{f} \otimes \matId + \matId \otimes \hat{f} -2 \, f_0 \otimes h_0 + 2 \, h_0 \otimes f_0.
  \end{aligned}
\end{equation}
The remarkable fact is now that, in the evaluation representation, thanks to the special relation (\ref{upper}) between $u$ and the energy eigenvalue, one finds that the shift (\ref{shift}) sets
\begin{equation}
\label{checkc}
\hat{e} = \hat{f}=0.
\end{equation}
In other words, the new evaluation parameter $\hat{u} \in \mathbb{C}$, pertaining to the first realisation, is simply
\begin{equation}
\hat{u} = 0.
\end{equation}
Having all the higher Yangian charges to be identically zero does not mean that the action on two-particle states is equally trivial for all of those, since the coproducts have a non-trivial tail. It simply means that, considering  the level-1 generators, the purely bilocal expressions
\begin{eqnarray}
\label{this}
e_0 \otimes h_0 - h_0 \otimes e_0 \qquad \mbox{and} \qquad f_0 \otimes h_0 - h_0 \otimes f_0
\end{eqnarray}
are themselves symmetries of the R-matrix, as can be directly verified. Moreover, one can check that (\ref{checkc}) is compatible with crossing symmetry, by applying the antipode relations reported in the previous section. 

\medskip

{\it $\bullet$ Secret symmetry}

\medskip

The first realisation is also very convenient to formulate the {\it secret} symmetry \cite{secret,Hoare:2014kmaa}:
\begin{equation}
\label{secret}
\Delta\big(\, \hat{b}\, \big) = \hat{b} \otimes \matId + \matId \otimes \hat{b} + e_0 \otimes f_0 + f_0 \otimes e_0,
\end{equation}
which, as in the case of $AdS_5$, has no counterpart at level 0. The representation\footnote{Since $\hat{b}$ never appears on the r.h.s. of any (anti-)commutation relation, there is always the freedom of adding terms proportional to the identity both to the generator and to the coproduct, without spoiling any relation - see also \cite{Borsato:2017lpf}. Such terms can be adjusted to ensure that the crossing symmetry is satisfied for the secret symmetry as well.} of $\hat{b}$  is given by
\begin{equation}
\label{ni}
\hat{b} = {\cal{B}}(\theta) \, (\bE_{11} - \bE_{22}).
\end{equation}
What we find here is that, in order to have a symmetry of the R-matrix, we must again set
\begin{equation}
\label{compa}
{\cal{B}}(\theta) = 0.
\end{equation} 
This means that the bilocal expression
\begin{equation}
\label{thats}
\Delta\big(\, \hat{b}\, \big) = e_0 \otimes f_0 + f_0 \otimes e_0
\end{equation}
is itself a symmetry of the R-matrix. Invariance of the R-matrix under (\ref{thats}) can also be checked directly.

\subsection{\label{sec:Scale}Scale invariance}

We begin by remarking that, by explicit check, all the results of the previous two sections equally apply both to solution 3 and to solution 5 in the terminology of \cite{Fontanella:2017rvu}, which we remind the reader were associated to two inequivalent limits one could take from the massive S-matrix, and also to an asymptotic limit of the two distinct ${\cal{N}}=1$ solutions of Fendley's \cite{Fendley:1990cy}. In each of these two cases, all our results exactly hold for both right and left movers. Quite interestingly, they also apply more generally, and in the exact same form, to the two original Fendley's solutions (also reported in \cite{Fontanella:2017rvu}).  

We speculate that this might be related to the inherent scale-invariance of the problem, combined with supersymmetry. According to Zamolodchikov \cite{Zamol2}, the right-right and left-left massless scattering matrices are in a sense to be regarded as {\it conformal} objects. In the picture of massless scattering as an interpolating flow between CFTs, these two S-matrices are preserved through the renormalisation group flow and remain unchanged all the way to the UV and IR critical points, where they characterise the CFT. What normally instead {\it drives} the flow is the mixed right-left scattering, which typically bears memory of the surviving mass-scale in the problem. For instance, the respective central charges of the UV and IR CFTs can be computed using these massless S-matrices and their associated thermodynamic Bethe ans\"atze (as will be described in much detail in \cite{Diego}). 

In our situation, the mixed right-left scattering is trivial, and the two right-right and left-left S-matrices must already describe a scale-invariant situation, where all mass-scales have disappeared. This can be seen by the fact that a rescaling of the supercharges (akin to a change of units of measurement for the energy)
\begin{equation}
\mathfrak{Q} \to M \mathfrak{Q}, \qquad \mathfrak{G} \to M \mathfrak{G}, 
\end{equation} 
amounts to a shift in rapidities
\begin{equation}
\theta_1 \to \theta_1 + 2 \log M, \qquad \theta_2 \to \theta_2 + 2 \log M,
\end{equation}
which is of course completely inconsequential for the S-matrix\footnote{We thank Ingo Runkel for pointing this out.}. This is reminiscent of the situation in $SU(2)_1$ CFTs \cite{Zamol2}. Quantum group structures in CFT are the subject of \cite{BLZ}, see also \cite{MacKay:2004tc} - esp. section 3.5 - and \cite{Bernard:1991vq}.

We speculate that the particular Yangian representation we have found might also be a sign of scale invariance. The Yangian evaluation parameter is traditionally hiding an $\hbar$ parameter, which introduces a surreptitious scale in the problem. This parameter can be rescaled away by redefining the Yangian generators. Nevertheless, in considerations of specific models, one can use still this parameter to tune a mass-scale. However, this scale is absent when the spectral parameter vanishes. In a sense, it is as if the mass scale set by $\hbar$ had been taken to infinity.

To corroborate this viewpoint, we notice that the particular type of Yangian (\ref{this}) and (\ref{thats}) seems to coincide with what is expected from the Yangian superalgebras of local and non-local charges \cite{Bernard:1990ys} appearing as symmetries of specific conformal field theories. In fact, working in Drinfeld's first realisation and studying supergroup sigma models, \cite{Creutzig:2010hr} finds that the Yangian Serre relations are only valid for the subset of fields transforming in representations which can be {\it trivially lifted} to representations of the Yangian. A {\it trivial lift} is precisely the one we have, when we find $\hat{J}=0$ for all the generators in Drinfeld's first realisation. A similar phenomenon seems to feature in \cite{Corn:2010uj}. 

This is also one situation where imposing a constraint between the spectral parameter and the energy does not break the difference form of the S-matrix - see also the discussion in \cite{Mitev:2012vt}.

\section{\label{sec:Massive}Application to the massive case}
In this section, we apply the procedure based on the inversion relation, which we have used for the massless S-matrix, to the massive case, since also the massive S-matrix satisfies the free-fermion condition. We shall not be able to proceed to the same extent as in the massless case, given the substantial complication of working with the massive S-matrix. Nevertheless, we will find that a series of rather miraculous occurrences, cancellations and simplifications, makes it possible to reach a remarkably advanced stage in the process. We shall still be able to reduce the problem to a factorisation condition, and to write auxiliary Bethe equations for the potential zeroes of the transfer matrix. These will however simply be too complicated to attack at this stage, and, unfortunately, even to compare to the proposal of \cite{Sorokin:2011rr}.  

We begin by writing the massive S-matrix in the notation of section \ref{sec:Aux}. We will refer to \cite{Hoare:2014kma} for the detailed expression of the S-matrix entries, which we shall not report here for the sake of a lighter presentation. We will use the exact same terminology of \cite{Hoare:2014kma} - cf. eq.s (3.2)-(3.6) of that paper -, setting $\alpha = \alpha' = 1$ for reasons of simplicity. We will also disregard the overall dressing factor, irrelevant to all the points we will manage to make in this section. One has   
\begin{eqnarray}
&&a_+ = 1, \qquad a_- = - \frac{S_2}{S_1}, \qquad b_+ = \frac{T_1}{S_1}, \qquad b_- = \frac{T_2}{S_1},\nonumber\\
&&c_+ = \frac{R_1}{S_1}, \qquad c_- = \frac{R_1}{S_1}, \qquad d_+ = \frac{Q_1}{S_1}, \qquad b_- = - \frac{Q_1}{S_1},
\end{eqnarray}
such that
\begin{eqnarray}
a_+ a_- + b_+ b_- = c_+ c_- + d_+ d_-.
\end{eqnarray}
We choose again the associated S-matrix $S^{(1)}$ to be defined by the map 
\begin{eqnarray}
&&a_\pm \to a^{(1)}_\pm = - b_\pm, \qquad b_\pm \to b^{(1)}_\pm = a_\pm, \nonumber\\
&&c_\pm \to c^{(1)}_\pm = c_\pm, \qquad\quad d_\pm \to d^{(1)}_\pm = - d_\pm.
\end{eqnarray}    
The S-matrix $S^{(1)}$ still clearly satisfies the free-fermion condition.

The transformation $X$ in (\ref{x}) and (\ref{suchthatx}) does not work for the massive case, nor can one find a {\it constant} matrix which produces an upper triangular form like (\ref{suchthatx}) for computing $T T^{(1)}$. This could invalidate the whole procedure, were it not for the following fortunate fact. A matrix $X$ exists, which does the job, and {\it only depends on the variables $x^\pm_0$} of the auxiliary space, and not on the variables $x_i^\pm$, $i=1,...,N$, of the quantum space. This is of course good enough for the similarity transformation to cancel out telescopically, and for the procedure to continue. One can take for instance  
\begin{eqnarray}
\label{xx}
X = \frac{1}{\sqrt{2}} \begin{pmatrix}0&1&1&0\\1&0&0&1\\1&0&0&-\mu\\0&1&-1&0\end{pmatrix},
\end{eqnarray}
with
\begin{eqnarray}
\label{virt}
\mu = -\xi + \sqrt{\xi^2 +1}, \qquad \xi = \frac{b_-^2 + c_+^2 - d_-^2 - 1}{2 \, c_+ d_-} = \frac{1}{2} \frac{\Big[\sqrt{x_0^-} + \sqrt{x_0^+}\, \Big]\big(x_0^+ \, x_0^- \, -1\big)}{x_0^- \sqrt{x_0^+} - x_0^+ \sqrt{x_0^-}} = \frac{i \, m_0}{4 h} \, \frac{1}{\sin^2 \frac{p_0}{4}},
\end{eqnarray}
where $m_0$ is the mass of the auxiliary particle, and $p_0$ its momentum, such that $\frac{x^+_0}{x_0^-} = e^{i p_0}$. For long representations like the massive ones, the mass is a free unconstrained parameter, which might even depend itself on the momentum and on the coupling constant $h$ \cite{Hoare:2014kma}. Notice that, if $m_0=0$, then $\xi$ vanishes and $\mu = 1$, which reduces the transformation $X$ to the one we utilised for the massless case. 

The upper triangular form one obtains for computing $T T^{(1)}$ is now characterised by diagonal entries
\begin{eqnarray}
m_+ = \tau_+ \, \mathfrak{1} = \big(c_+^2 - b_+ b_-\big) \mathfrak{1}, \qquad m_- = \tau_- \, \mathfrak{1} = \big(d_-^2 - b_+ b_-\big) \mathfrak{1}, \qquad n_\pm = \rho_\pm \, \sigma_3 = \big(- b_+ \mp c_+ d_- \mu^{\mp 1}\big) \sigma_3.\nonumber
\end{eqnarray}
All the quantities depend on the auxiliary variables $x_0^\pm$ {\it and} on the variables of the specific $i$-th particle, {\it i.e.} $x_i^\pm$. This is with the exception of $\mu$, which, as we pointed out, only depends on $x_0^\pm$.
  
One therefore gets
\begin{eqnarray}
\label{frommo}
T T^{(1)} = \prod_{i=1}^N \tau_+(x_0^\pm, x_i^\pm) +  \prod_{i=1}^N \tau_-(x_0^\pm, x_i^\pm) +\Bigg[\prod_{i=1}^N \rho_+(x_0^\pm, x_i^\pm)+\prod_{i=1}^N \rho_-(x_0^\pm, x_i^\pm)\Bigg]F\ ,
\end{eqnarray}
where $F$ denotes again the fermionic number of the eigenstate. Let us notice that, by virtue of (\ref{virt}), the entire term proportional to $F$ in (\ref{frommo}) is a polynomial in $b_\pm$, $c_+$, $d_-$ and $\xi$ for any $N$. We have performed some explicit checks, for up to $N=2$, that (\ref{frommo}) is correct.

We now need to be more precise on the relation between $S$ and $S^{(1)}$. One can check that
\begin{eqnarray}
\label{relaz}
S_{0'i}^{(1)}(x_0^\pm, x_i^\pm) = a_-(x_0^\pm, x_i^\pm) \, G \, S_{0'i}\bigg[\frac{1}{x_0^\pm}, x_i^\pm\bigg] G^{-1}\ , \qquad G = \mathsf{E}_{12} + i \, \mathsf{E}_{21}\ , \qquad i=1,..,N,
\end{eqnarray}
where the similarity transformation is meant to be performed in the auxiliary space $0'$. This shows that the S-matrix $S^{(1)}$ satisfies the Yang-Baxter equation.
We therefore have
\begin{eqnarray}
\label{that}
T^{(1)} (x_0^\pm, x_i^\pm) = \Bigg[\prod_{i=1}^N a_-(x_0^\pm, x_i^\pm)\Bigg] \times T\bigg(\frac{1}{x_0^\pm}, x_i^\pm\bigg), 
\end{eqnarray}
which is again nothing else than the crossing transformation on the auxiliary space, up to an overall scalar factor. Notice that, in the massless relativistic limit, $a_- \to 1$ and we recover exactly formula (5.2) in  \cite{Fontanella:2017rvu}. We recognise in (\ref{frommo}) a factorisation problem akin to solving a crossing equation. This was difficult in the massless case, now it looks extremely daunting.

We can still try to identify the potential zeroes of the transfer-matrix eigenvalues, in order to get the auxiliary Bethe equations. We split 
\begin{eqnarray}
\label{times}
\notag
&&T T^{(1)} = \bigg[\prod_{i=1}^N A_1(x_0^\pm, x_i^\pm) + F \bigg]\times \bigg[\prod_{i=1}^N A_3(x_0^\pm, x_i^\pm) + F \prod_{i=1}^N A_4(x_0^\pm, x_i^\pm)\bigg],
\end{eqnarray}
where 
\begin{eqnarray}
A_1(x_0^\pm, x_i^\pm) = \prod_{i=1}^N \frac{\tau_+(x_0^\pm, x_i^\pm)}{\rho_+(x_0^\pm, x_i^\pm)}, \qquad A_3 =  \prod_{i=1}^N \rho_-(x_0^\pm, x_i^\pm), \qquad A_4 = \prod_{i=1}^N \rho_+(x_0^\pm, x_i^\pm).
\end{eqnarray}
The fact that one can do this is again a rather miraculous occurrence, thanks to the identity
\begin{eqnarray}
\label{thanks}
\frac{\tau_+(x_0^\pm, x_i^\pm) \, \tau_-(x_0^\pm, x_i^\pm)}{\rho_+(x_0^\pm, x_i^\pm) \, \rho_-(x_0^\pm, x_i^\pm)} = 1, \qquad \forall \, \, \, i=1,...,N.
\end{eqnarray}
One therefore sees that potential zeroes of the transfer matrix can come from either of the two conditions
\begin{eqnarray}
\label{au}
\prod_{i=1}^N \frac{\tau_+(x_0^\pm, x_i^\pm)}{\rho_+(x_0^\pm, x_i^\pm)} = - F, \qquad \prod_{i=1}^N \frac{\rho_-(x_0^\pm, x_i^\pm)}{\rho_+(x_0^\pm, x_i^\pm)}= - F.
\end{eqnarray}
In fact, quite remarkably, one can prove that each of two auxiliary Bethe equations (\ref{au}) maps into itself under crossing\footnote{We observe that the quantity $\xi$ is crossing-invariant (therefore so is $\mu$).}, since
\begin{eqnarray}
\label{crox}
\tau_\pm \bigg(\frac{1}{x_0^\pm}, x_i^\pm\bigg) = \frac{\tau_\pm (x_0^\pm, x_i^\pm)}{a_-^2 (x_0^\pm, x_i^\pm)}, \qquad \rho_\pm \bigg(\frac{1}{x_0^\pm}, x_i^\pm\bigg) = \frac{\rho_\pm (x_0^\pm, x_i^\pm)}{a_-^2 (x_0^\pm, x_i^\pm)}, \qquad i=1,...,N,
\end{eqnarray}
and of course $\frac{1}{F} = F$.
Eq. (\ref{crox}) is consistent with the property
of $T T^{(1)}$ under crossing, given that 
\begin{eqnarray}
a_- \bigg(\frac{1}{x_0^\pm}, x_i^\pm\bigg) = \frac{1}{a_- (x_0^\pm, x_i^\pm)}.
\end{eqnarray}
Interestingly enough, this situation is very similar to the one described in \cite{Fontanella:2017rvu}, when discussing the Bethe ansatz for Fendley's S-matrix \cite{Fendley:1990cy} of relativistic ${\cal{N}}=1$ theories.

The auxiliary Bethe equations (\ref{au}), although more than we could initially hope for, are still too complicated to easily manage. We have not been able to simplify their explicit expressions to even attempt comparison with \cite{Sorokin:2011rr}. What we have been able to do is to check that in the massless limit, by carefully resolving the expected $\frac{0}{0}$ expressions coming from the function $f$ as in \cite{Hoare:2014kma,Fontanella:2017rvu}, the second equation (\ref{au}) reduces exactly to (\ref{just}), while the first equation is never satisfied in the limit, which reassures us that all the zeroes can {\it only} come from (\ref{just}).

\section{\label{sec:Concl}Conclusions}

The focus of this paper is the integrable superstring theory on $AdS_2 \times S^2 \times T^6$. While massive magnon representations are long, the massless ones are short, therefore it seems worthwhile concentrating on them first, and trying to reach a complete understanding of their dynamics. Integrable scattering of massless excitations is traditionally a very powerful tool to describe the non-perturbative behaviour of the theory, and we hope that this shall be the case for the $AdS_2/CFT_1$ correspondence as well.

To move steps in this direction, we have here explicitly computed the transfer-matrix eigenvalues starting from the S-matrix of integrable $AdS_2 \times S^2 \times T^6$ superstrings, up to 5 particles. At that stage, we have conjectured the general behaviour, based on the knowledge of the location of the potential zeroes determined in \cite{Fontanella:2017rvu}. We have then used the conjectured form of the eigenvalues to write down a set of massless Bethe ansatz equations. The same procedure applies to the relativistic and to the non-relativistic situation. However, it is only in the relativistic case that the right and left modes decouple. Following Zamolodchikov, this situation should describe a critical fixed point, hence we are brought to conjecture that the relativistic massless Bethe ansatz we obtain might capture the integrable structure of some limiting supersymmetric {\it worldsheet} CFT. As anticipated in \cite{Borsato:2016xns} in the $AdS_3$ context, and as it will be extensively discussed in that framework in \cite{Diego}, the relativistic scattering theory one finds\footnote{From this particular point of view, the arguments applicable to $AdS_3$ can to a large extent be applied to $AdS_2$ as well.} is completely non-perturbative, and it is not clear how to describe it (or test it) from the worldsheet sigma-model perspective\footnote{We thank Diego Bombardelli and Bogdan Stefa\'nski for many clarifying discussions about this point.}.

The most urgent task is now to subject the Bethe ansatz we propose to a thorough list of tests of internal consistency. Among the checks that need to be setup, one should find a way to address the issues of frame-dependence, the extraction of the global charges, and how to develop an efficient method of solution. We also need to determine whether our Bethe ansatz would follow from a worldsheet analysis, and whether the method we are using truly captures enough information to reconstruct the complete massless spectrum. 

Subsequently, one should derive the thermodynamic Bethe ansatz (TBA) equations. One should then solve them in order to find the central charge \cite{Zamol2}, which could pin down which 2D critical theory our relativistic massless scattering theory is describing \cite{Diego}. We would then be ready to consider our results in the light of \cite{Hofman:2011zj} for example, and address the issues of scale and conformal invariance in our S-matrix approach against the expectations from standard $AdS_2$ holographic arguments. More specifically, \cite{Diego} will show how this paradigm can be applied to the analogous situation in the context of the $AdS_3 \times S^3 \times T^4$ superstring. The same relativistic limit, which we have here applied to the $AdS_2$ massless sector, produces in $AdS_3$ a more standard Bethe ansatz. The associated (reduced set of) TBA equations can be written down, in a closed and tractable form. One of the main results presented in \cite{Diego} will be to show how a central charge for the putative 2D CFT can be determined unambiguously from these TBA equations, notwithstanding the notorius difficulties in dealing with massless TBAs. Such TBAs are in fact hard to analyse within a perturbative scheme, as the contribution from massless particles is not sufficiently suppressed, see also \cite{MI}. In the $AdS_2$ case, although we have not reported them here, we have managed to write down the TBA equations, based on the asymptotic Bethe ansatz conjectured in this paper. However, the kernels one encounters at the very final stage of the computation are wilder and, in particular, do not seem to have a definite parity $\theta \to - \theta$, which turns out to be a surprisingly tough obstacle in obtaining a closed form for the vacuum energy. For this technical reason, we are still unable to unambiguously fix the central charge, and we might have to eventually resort to a numerical approach if all else fails.    

The results we have obtained for the massive S-matrix, although still rather difficult to tame, give us the first glimpse of hope that the procedure relying on the free-fermion condition might concretely work in that case as well. One hopes to make contact with the proposed Bethe ansatz of \cite{Sorokin:2011rr} eventually, although we are not in that position yet. The feature of the massive representations of being long might set them in a separate category from the massless ones, and the role they play remains to be fully understood\footnote{We thank B. Hoare for discussions about this point}. We hope that our findings will nevertheless give a little contribution to that understanding, and, in general, to the discussion concerning superstrings in $AdS_2$ and their CFT dual description.

\section*{\label{sec:Ackn}Acknowledgments}

We thank Diego Bombardelli, Andrea Fontanella, Ben Hoare and Bogdan Stefa\'nski for the very many inspiring discussions and illuminating explanations, and for very useful comments on the manuscript. We thank Romuald Janik, Olof Ohlsson Sax, Roberto Tateo and Benoit Vicedo for very helpful discussions. We thank Micheal Abbott, In$\hat{\mbox{e}}$s Aniceto, Gleb Arutyunov, Patrick Dorey, Davide Fioravanti, Tomek Lukowski, Antonio Pittelli, Andrea Prinsloo, Marco Rossi, Ingo Runkel and Ryo Suzuki for insightful discussions. We thank the STFC for support under the Consolidated Grant project nr. ST/L000490/1. We thank the Galileo Galilei Institute for Theoretical Physics (GGI) for the hospitality and INFN for partial support within the program {\it New Developments in $AdS_3/CFT_2$ Holography}.

\section*{\label{sec:Data}Data management}

No data beyond those presented in this paper are needed to validate its results.

\end{document}